# Electrostatics in dissipative particle dynamics using Ewald sums with point charges

**Ketzasmin A. Terrón-Mejía, Roberto López-Rendón**

Laboratorio de Bioingeniería Molecular a Multiescala, Facultad de Ciencias,
Universidad Autónoma del Estado de México,
Av. Instituto Literario 100, Toluca 50000, Estado de México, Mexico

**Armando Gama Goicochea**[*]
División de Ingeniería Química y Bioquímica, Tecnológico de Estudios Superiores de Ecatepec, Av. Tecnológico s/n, Ecatepec 55210, Estado de México, Mexico

## Abstract

A proper treatment of electrostatic interactions is crucial for the accurate calculation of forces in computer simulations. Electrostatic interactions are typically modeled using Ewald – based methods, which have become one of the cornerstones upon which many other methods for the numerical computation of electrostatic interactions are based. However, their use with charge distributions rather than point charges requires the inclusion of ansatz for the solutions of the Poisson equation, since there is no exact solution known for smeared out charges. The interest for incorporating electrostatic interactions at the scales of length and time that are relevant for the study the physics of soft condensed matter has increased considerably. Using mesoscale simulation techniques, such as dissipative particle dynamics (DPD), allows us to reach longer time scales in numerical simulations, without abandoning the particulate description of the problem. The main problem with incorporating electrostatics into DPD simulations is that DPD particles are soft and those particles with opposite charge can form artificial clusters of ions. Here we show that one can incorporate the electrostatic interactions through Ewald sums with point charges in DPD if larger values of coarse – graining degree are used, where DPD is truly mesoscopic. Using point charges with larger excluded volume interactions the artificial formation of ionic pairs with point charges can be avoided, and one obtains correct predictions. We establish ranges of parameters useful for detecting boundaries where artificial formation of ionic pairs occurs. Lastly, using point charges we predict the scaling properties of polyelectrolytes in solvents of varying quality and obtain predictions that are in agreement with calculations that use other methods, and with recent experimental results.

---

[*] Corresponding author. Electronic mail: agama@alumni.stanford.edu

# I. INTRODUCTION

One of the greatest current challenges in theoretical physics is to understand the basic principles that govern soft condensed matter systems. Electrostatic interactions between charges are the leading cause of many phenomena that occur in the fields of condensed matter physics, chemistry, molecular biology and materials science. Among the various components of molecular interactions, electrostatic interactions deserve special attention because of their long range, strength, and significant influence on charged and/or polar molecules, including water [1], aqueous solutions [2], and amino acids or nucleic acids [3]. Electrostatics plays a major role in defining the mechanisms of protein – protein complex formation, molecular recognitions, thermal stability, conformational adaptabilities and protein movements. A wide variety of theoretical approaches, ranging from quantum mechanical ab initio methods, classic Maxwell's theory of electromagnetism, generalized Born algorithms, to phenomenological modifications of Coulomb's law, have been used for electrostatic analysis [4]. Quantum mechanical methods are generally accurate, but are very demanding computationally to be used for large chemical and biological systems. Generalized Born algorithms are fast, but depend on other methods for calibrations. The Poisson equation, derived from Gauss's law and using linear polarization, provides a relatively simple and accurate method while being a much less expensive description of electrostatic interactions for a given charge source density [5].

Electrostatics poses several challenges for the mesoscopic modeling approach: first, the $1/r$ potential is long – ranged and cannot be simply integrated away. This is also well – known from the standard virial expansion: if Coulombic interactions are included, the expansion does not converge [6]. These issues have led to different approaches, including the implicit inclusion of charges via iterative coarse-graining procedures, and the explicit inclusion of charges and attempts to include some aspects of Coulomb interactions [7-8]. In a recent review by Zhang *et al*. [9] on the role of electrostatics in protein – protein interactions, documented how some biological complexes are formed of identical macromolecules (homo-complexes), while others involve different entities (hetero – complexes). The main difference between these two cases is the net charge of the monomers, which for homo-dimers is the same as for both monomers, while for hetero – complexes the monomers frequently carry opposite net charges. Such a difference is expected to result in different roles of electrostatics on the protein – protein recognition at very large distance, at which the distribution of the charges is not important but the net charge is. On the other hand, Raval *et al*. [10] performed a comprehensive study reaching at least 100 µs using molecular dynamics simulations for 24 proteins. For most systems, the structures drift away from the native state, even when starting from the experimental structure. They concluded that this is most likely a limitation of the available point – charge force fields. Therefore, the treatment of long-range electrostatic interactions must be performed with great care.

Computationally, electrostatic interactions are calculated usually through lattice – sum methods, such as the Ewald summation [11] technique, the particle – particle – particle – mesh ($P^3M$) [12], or particle – mesh Ewald PME [13] methods. It is well – known that a simple real – space summation of interactions between points within a cutoff – radius does slowly and only conditionally converge in the limit of large cutoffs. The Ewald summation is a standard technique to compute the electrostatic interactions with properly converging sums. The main idea behind the Ewald summation is to split up the $1/r$ dependence of the Coulomb interaction into an exponentially decaying part and a residual potential. The exponentially decaying potential is typically that of a unit point charge compensated by a Gaussian charge cloud with the opposite charge. The residual potential is that of a Gaussian charge cloud with the same width and unit charge. At present there exist several mesh implementations of the Ewald sum — similar in spirit but different in detail. A detailed numerical analysis of several Ewald mesh methods is available in reference [14].

The application and development of the Ewald method has been very popular in molecular dynamics simulations (MD), which require that the equations of motion be solved with a time scale of the order of femtoseconds and a length scale of the order of angstroms. This might be the most accurate classical approach for studying mesoscale phenomena, although it is still time consuming. One alternative approach to study mesoscale phenomena is the method known as dissipative particle dynamics (DPD). The model, initially proposed by Hoogerbrugge and Koelman [15], consists of reducing the complexity of the atomistic degrees of freedom of the system through the use of a coarse – grained approach. This reduction makes the DPD method considerably faster compared with MD. There are several works that include electrostatics into the DPD model. The main problem with using the Ewald method in DPD simulations is that the DPD particles are soft, and those particles with opposite charge can overlap and form artificial clusters of ions, with exceedingly large electrostatic interaction. To get around this problem, the first approach, due to Groot [16], follows the idea of regularizing the electrostatic interaction at short distances, which can be done by smearing out the charges within the DPD particles. When the separation of oppositely charged particles is large, the potential has the usual $1/r$ form but there must be a penalty at distances shorter than some cutoff that defines the radius over which the charge is smeared out. Then, the electrostatic field is solved on a grid as in the $P^3M$ method [16]. Another method is that of González – Melchor *et al*. [17], which follows closely Groot's idea for the smearing of the charges on the DPD particles, but solves the electrostatics using the standard Ewald summation, instead of a grid solution. This method allows one to use a standard approach to calculate the electrostatic energy and the force between charge density distributions. Although the inclusion of charge density distributions does not directly increase the computational cost in the Ewald summation itself, this method becomes computationally more demanding than the one adopted by Groot as the system's size is increased. Except for the latter feature, these two approaches are essentially equivalent [18] and good agreement has been found between them in the radial distribution functions of charged particles in bulk electrolytes and polyelectrolyte – surfactant solutions. On the other hand, Praprotnik *et al*. [19] have developed a very different multiscale method from those described previously. In their approach, termed adaptive resolution, the spatial resolution of the system is tuned adaptively. Further contributions to the implementation of electrostatic interactions in DPD simulations are found in the Refs [20, 21]. For a recent review on the state of the art of the use of electrostatic interactions in the computational modeling of complex systems, see reference [22].

Although the use of Ewald sums is the most popular method for the handling of electrostatic interactions in computational simulations because of its relative simplicity, their use with charge distributions rather than point charges requires the inclusion of ansatz for the solutions of the Poisson equation, since there is no exact solution known for smeared out charges [23]. This approach is not completely satisfactory since it does not lend itself easily to generalization for applications to different coarse graining degrees. In reviewing the original assumptions that led to the use of charge distributions in DPD [16], we have noted that the reason why artificial ionic pairs could form is that the coarse graining degree – the number of water molecules grouped into a DPD bead – used was the minimum possible, namely one water molecule per DPD particle. In such case the (non – electrostatic) DPD particle – particle interaction allows for considerable overlap between the particles, which in turn could allow for the almost complete overlap of their electric charges, leading to a discontinuity of the Coulomb interaction. However, the DPD model like most coarse – grained models works best when the coarse graining degree is large, which means that the particle – particle interactions are strong enough to prevent the complete penetration of the DPD particles, allowing one to use point charges, for which the solution of Poisson's equation is known. To prove such approach is the purpose of the present work, which to our knowledge is a topic that to date has not yet been explored. The remainder of this paper is organized as follows. The general equations of the DPD model, and the basic theoretical framework of electrostatic interactions with Ewald sums are presented in Section II. The details of the simulations are presented in Section III. In Section IV, our results and their discussion are presented. Finally, the conclusions are drawn in Section V.

## II. THEORETICAL FRAMEWORK

**A. The DPD model**

In the DPD model, the particles do not represent actual molecules but rather groups of atoms or molecules that interact through simple, linearly decaying force laws, that is why it is a mesoscopic approach. The theoretical foundations of the DPD method can be found in various sources [24 – 27], therefore we shall only outline some general aspects of this technique, for brevity. DPD simulations are similar to traditional molecular dynamics algorithms [28] in that one must integrate Newton's second law of motion using finite time steps to obtain the particles' positions and momenta from the total force

$$\frac{d\mathbf{r}_i}{dt} = \mathbf{v}_i, \qquad m_i \frac{d\mathbf{v}_i}{dt} = \mathbf{F}_i, \tag{1}$$

where $\mathbf{r}_i$, $\mathbf{v}_i$ and $\mathbf{F}_i$ denote the position, velocity, and the total force acting on particle $i$, respectively. A difference from MD is that the total force in the DPD model involves three different pairwise additive forces acting between any two particles $i$ and $j$, placed a distance $r_{ij}$ apart, namely, conservative force ($\mathbf{F}^C$), random ($\mathbf{F}^R$), and dissipative ($\mathbf{F}^D$), components. In its traditional form, the DPD total force is the sum of these three components [24],

$$\mathbf{F}_{ij} = \sum_{i \neq j}^{N}[\mathbf{F}^C(\mathbf{r}_{ij}) + \mathbf{F}^R(\mathbf{r}_{ij}) + \mathbf{F}^D(\mathbf{r}_{ij})] \ . \tag{2}$$

The conservative force determines the thermodynamics of the DPD system and is defined by a soft repulsion, representing the excluded volume interactions:

$$\mathbf{F}_{ij}^C = \begin{cases} a_{ij}(1 - r_{ij})\hat{\mathbf{r}}_{ij} & r_{ij} \leq R_c \\ 0 & r_{ij} > R_c \end{cases} \tag{3}$$

where $a_{ij}$ is the maximum repulsion strength between a pair of particles $i$ and $j$, and $\mathbf{r}_{ij} = \mathbf{r}_i - \mathbf{r}_j$, $r_{ij} = |\mathbf{r}_{ij}|$, $\hat{\mathbf{r}}_{ij} = \mathbf{r}_{ij}/r_{ij}$. The dissipative and the random forces are given by

$$\mathbf{F}_{ij}^D = -\gamma \omega^D(r_{ij})[\hat{\mathbf{r}}_{ij} \cdot \mathbf{v}_{ij}]\hat{\mathbf{r}}_{ij} \tag{4}$$

$$\mathbf{F}_{ij}^R = \sigma \omega^R(r_{ij})\xi_{ij}\hat{\mathbf{r}}_{ij} \tag{5}$$

where $\gamma$ is the dissipation streng, $\sigma$ is the noise amplitude, $\omega^D$ and $\omega^R$ are distance dependent weight functions, $\mathbf{v}_{ij} = \mathbf{v}_i - \mathbf{v}_j$ is the relative velocity between the particles, and $\xi_{ij} = \xi_{ji}$ is a random number distributed between 0 and 1 with Gaussian distribution and variance, $1/\Delta t$ where $\Delta t$ is the time step of the simulation. The magnitude of the dissipative and stochastic forces are related through the fluctuation-dissipation theorem [27]

$$\omega^D(r_{ij}) = [\omega^R(r_{ij})]^2 = max\left\{\left(1 - \frac{r_{ij}}{R_c}\right)^2, 0\right\} \tag{6}$$

where $R_c$ is a cut-off distance. At interparticle distances larger than $R_c$, all forces are equal to zero. This simple distance dependence of the forces, which is a good approximation to the one obtained by spatially averaging a van der Waals - type interaction, allows one to use relatively large integration time steps. The strengths of the dissipative and random forces are related in a way that keeps the temperature ($T$) internally fixed, $k_B T = \frac{\sigma^2}{2\gamma}$; $k_B$ being Boltzmann's constant. The natural probability distribution function of the DPD model is that of the canonical ensemble, where $N$ (the total particle number), $V$ (volume), and $T$ are kept constant. All forces between particles $i$ and $j$ are zero beyond a finite cutoff radius $R_c$, which represents the intrinsic length scale of the DPD model and is usually also chosen as $R_c \equiv 1$. For more details about recent successful applications of the DPD model, the reader is referred to reference [29].

**B. Electrostatic interactions in DPD.**

In this section we start by presenting the standard Ewald sums as they are typically used in numerical simulations [28], followed by their specific implementation in DPD simulations. Let us first consider a periodic system of $N$ point charges $\{q_i\}$ located at position $\{\mathbf{r}_i\}$, $i = 1, \ldots, N$. The classical electrostatic energy of this system is given according to Coulomb's law by

$$U(\mathbf{r}^N) = \frac{1}{4\pi\varepsilon_0 \varepsilon_r} \frac{1}{2} \sum_{\mathbf{n}}^{*} \sum_{i=1}^{N} \sum_{j=1}^{N} \frac{q_i q_j}{|\mathbf{r}_{ij} + \mathbf{n}|}, \quad (7)$$

where $\mathbf{r}_{ij} = \mathbf{r}_i - \mathbf{r}_j$ and the summation over $\mathbf{n}$ is over all integer translations of the real space lattice vectors $\mathbf{n} = n_1 L_x \hat{\boldsymbol{\imath}} + n_2 L_y \hat{\boldsymbol{\jmath}} + n_3 L_z \widehat{\boldsymbol{k}}$ for integers $n_k$ ($k = 1, 2, 3$), and the asterisk indicates that the summation excludes all pairs $i = j$ when $\mathbf{n} = 0$, in the central simulation cell. The symbols $L_x$, $L_y$ and $L_z$ represent the sides of the simulation cell. The variables $\varepsilon_0$ and $\varepsilon_r$ are the permittivity of vacuum and the dielectric constant of water at room temperature, respectively. The summation in eq. (7) is not convergent unless the total charge of the system equals zero. The sum over **n** takes into account the periodic images. The expression in eq. (7) converges very slowly under the conditions where it does so, and is not a practical means of computing electrostatic energies for periodic systems. It is then possible to decompose the long-range electrostatic interactions given by eq. (7) into real space and reciprocal space contributions, leading to a short-ranged sum, which can be written as [11, 30] :

$$U(\mathbf{r}^N) = \frac{1}{4\pi\varepsilon_0\varepsilon_r}\left(\sum_i \sum_{j>i} q_i q_j \frac{\text{erfc}(\alpha r_{ij})}{r_{ij}} + \frac{2\pi}{V}\sum_{\mathbf{k}\neq 0}^{\infty} Q(k) S(\mathbf{k}) S(-\mathbf{k}) - \frac{\alpha}{\sqrt{\pi}}\sum_i^N q_i^2\right). \quad (8)$$

In eq. (8) we have omitted a term, usually referred to as the "surface term", which is given by $(2\pi/3V)\left|\sum_{i=1}^N q_i r_i\right|^2$, where $V$ is the simulation cell's volume, see for example [28]. It is necessary when the sphere formed by the central unit cell and its periodically repeated cells are surrounded by a medium with a drastically different permittivity. However, in all the case studies presented here, and in many other applications, the medium surrounding the cells with the charges is the same as that in which the charges are dissolved, namely the solvent. It is for this reason there is no need to include such surface term here. The functions $Q(k)$ and $S(\mathbf{k})$ in eq. (8) are defined by the following expressions:

$$Q(k) = \frac{e^{-k^2/4\alpha^2}}{k^2}, \quad S(\mathbf{k}) = \sum_{i=1}^N q_i e^{i\mathbf{k}\cdot\mathbf{r}_i}, \quad \mathbf{k} = \frac{2\pi}{L}(m_x, m_y, m_z), \quad k = |\mathbf{k}|, \quad (9)$$

where $\alpha$ is the parameter that controls the contribution of the Coulomb interaction in real space, $k$ is the magnitude of the reciprocal vector $\mathbf{k}$, while $m_x$, $m_y$, $m_z$ are integer numbers, and $\text{erfc}(x)$ is the complementary error function. The symbol $\alpha$ is chosen so that only pair interactions in the central cell are needed to be considered in evaluating the first term of eq. (8). Equation (8) is a good approach to the $1/r$ term given by eq. (7), capturing the full long-range nature of electrostatic interactions.

The Ewald sums technique is the most frequently employed route for the calculation of the electrostatic interactions in atomistic scale simulations, as expressed by eq. (8) - in the sense of it being a method that breaks up the total contribution of the electrostatic interactions into two sums, one performed in configurational space and one in reciprocal space -, but its implementation into DPD simulations has the problem that, because the conservative interactions between DPD particles are soft, the particles with opposite charge could form artificial ionic pairs. If the particles had hard cores, as in the hard sphere model, there would not appear a singularity in the Coulomb interaction even if the particles formed an ionic pair. To avoid the divergence of the electrostatic potential for soft DPD particles, Groot [16] introduced a charge distribution given by (where $r < R_e$)

$$\rho(r) = \frac{3}{\pi R_e^3}(1 - r/R_e) . \tag{10}$$

In eq. (10), $R_e$ is a smearing radius. For $r > R_e$, $\rho(r) = 0$. To calculate the electrostatic forces, Groot followed the method of Beckers *et al*. [31], where the electrostatic field is solved on a lattice. The charge distributions were spread out over the lattice nodes and the long-range part of the interaction potential was calculated by solving the Poisson's equation in real space. It was stated that the method works efficiently if the grid size is equal to the particle size. Afterward, González – Melchor *et al*. [17] considered a Slater – type charge distribution on DPD particles of the form

$$\rho(r) = \frac{q}{\pi \lambda^3} e^{-2r/\lambda} , \tag{11}$$

where λ is the decay length of the charge. When the charges are smeared out, as in eq. (11), the electrostatic interaction arising from them cannot in general be obtained analytically and approximate expressions must be used [23]. Thus, the reduced electrostatic force between two charge distributions with valence $Z_i$ and $Z_j$ respectively, separated by a distance $r^* = r_{ij}/R_c$ is given by [23]:

$$F_E^*(r^*) = \Gamma \frac{Z_i Z_j}{4\pi r^{*2}}\{1 - [1 + 2\beta^* r^* (1 + \beta^* r^*)]e^{-2\beta^* r^*}\} ,\qquad(12)$$

where $\Gamma = e^2/k_B T \varepsilon_0 \varepsilon_r R_c$ and $\beta^* = R_c \beta$ with $\beta = 5/(8\lambda)$, where $\lambda$ is the decay length of the charge distribution; $e$ is the electron charge. The value $R_c = 6.46$ Å is used, as corresponds to a coarse graining degree equal to three water molecules associated to a DPD particle. This force is added to the DPD conservative force to give the total conservative force, which will determine the thermodynamic behavior of the system.

### III. SIMULATION DETAILS

We use reduced units throughout this work, where all masses are taken as $m = 1.0$ and the cutoff radius is $R_C = 1.0$. The value of the constants in the random and dissipative forces, σ and γ, are chosen as 3.0 and 4.5 respectively, $k_B T$ is taken equal to 1.0 and the time step is set at $\delta t = 0.03$. All our simulations are performed at constant density and temperature, i.e., using the canonical ensemble. Three sets of simulations are carried out; the first is designed to test the use of point charges within the DPD particles by comparing the radial distribution functions of a linear polyelectrolyte in solution, obtained with point charges, with those obtained using charge distributions with Ewald sums. The second set of simulations explores the influence of increasing the coarse – graining degree in the formation of ionic pairs using point charges, in a system that contains ions, their counter ions and solvent particles. In the third set of simulations, we apply the point – charge method to the prediction of a scaling exponent for the radius of gyration of polyelectrolytes dissolved in solvent of different quality.

The first system consists of a single, linear polyelectrolyte in solution with $N = 32$ beads, all of which carry a positive charge equal to $q = 0.4e$ either as point charges or as charge distributions, in addition to solvent particles and negative counter ions in a cubic simulation box whose volume is $V = 15.85 \times 15.85 \times 15.85$ DPD units, to fix the total density at $\rho = 3.0$. The beads are joined by freely rotating, harmonic springs with spring constant $k_r = 100.0$ and equilibrium length $r_0 = 0.7$. The DPD parameters of the interaction matrix are $a_{ii} = 78.0$ for particles of the same type, and $a_{ij} = 78.0$, for particles of different type, which correspond to theta solvent conditions, while for the Ewald's sums parameters, the force in real space is truncated at $R_{ew} = 3.5 R_C$ using $\alpha_{ew} = 0.9695$; for the reciprocal space we use the maximum vector $k_{max} = (5,5,5)$, and $\beta^* = 0.929$ for the charge distributions, see eq. 12. This choice of parameters for the Ewald sums is made following the choice of other groups who used distributions of charge with Ewald sums in the DPD model [16 – 18], so that our results can be compared with theirs. The value of $R_{ew}$ (3.5 $R_C$) was chosen so that the analytical approximation given by eq. (12) is indistinguishable from the force used in reference [16] using that re scaling. The values for $\alpha_{ew}$ and $k_{max}$ were chosen empirically for computational efficiency, and the value for $\beta^*$ was chosen to match the strength of the interaction of two charge distributions when their relative distance is zero using either the charge distribution given by eq. (10) or that given by eq. (11).

We perform also simulations at increasing values of the coarse – graining degree, i. e., the number of water molecules grouped into a DPD particle, while keeping the rest of the parameters fixed; this constitutes the second set of simulations. Such coarse – graining values are $N_m = 1, 2, 3, 4$ and $5$, which correspond to $a_{ij} = 25, 50, 78, 104$ and $130$, respectively; at every coarse – graining level we calculate the radial distribution functions at increasing values of the point charges in the interval $[0.0, 1.0]e$ with increments $\Delta q = 0.1e$. These systems are composed of 500 ions and 500 counter ions; the absolute magnitude of the charge is the same for ions and their counter ions, to which 2000 particles of solvent in a simulation box are added, in a volume $V = 10 \times 10 \times 10$. With the information obtained from these simulations we construct a diagram that separates stable mixtures from those where ionic pairs are formed.

The third set of simulations, referred to at the beginning of this section, is designed to apply the point – charge method to the calculation of the scaling exponent of the radius of gyration for linear polyelectrolytes using point charges, under different solvent conditions [32]. For polyelectrolytes under conditions of theta solvent, the DPD interaction parameters between (*p*)olyelectrolyte particles and (*s*)olvent particles are $a_{pp} = a_{ss} = a_{ps} = 78$. The polymerization degree of the linear chains is varied as follows: $N = 20, 40, 60, 80, 100, 120, 140$ and $160$; the lateral size of the cubic simulation boxes is $L = 13.572, 17.100, 19.574, 21.544, 23.208, 24.662, 25.962$ and $27.144$, respectively for each polymerization degree. These are highly charged polyelectrolytes as each monomer has a charge $q = 0.4e$. The parameters of the harmonic spring used to bond beads in the polymer chains are $k_r = 100.0$ (spring constant) and $r_0 = 0.7$ (equilibrium length) [33], as in the first set of simulations. A single polymer chain is modeled in these cases, which means the polymer concentration is $C_p = 0.008$; however, the salt concentration, $C_s$, is increased through the addition of monomeric ions, which also carry point charges. The salt concentrations modeled are $C_s = 0.060M, 0.300M, 0.910M, 1.220M, 1.530M$, and $1.840M$. For the model of salt, we use tetravalent salt where the ions have a charge $q = 0.4e$ and the counter ions have a charge $q = -0.1e$. For polyelectrolytes under good solvent conditions, the DPD interaction parameters are chosen as $a_{ps} = 39$, $a_{pp} = a_{ss} = 78$, while the rest of parameters are the same as for the theta solvent case. In all cases, simulations of 50 blocks of $10^5$ DPD time steps each are carried out, of which the first 10 blocks are used to reach equilibrium and the rest for the production phase. All calculations reported in this work have been performed using the SIMES code, which is designed to study complex systems at the mesoscopic scale using graphics processors technology (GPUs) [34].

## IV. RESULTS AND DISCUSSION

In Fig.1 we present a schematic illustration of the (a) charge distribution model, and (b) point – charge model within the DPD particles, for the particular case of coarse – graining degree equal to three. The total charge contained is the same for both cases shown; the maximum of the charge distribution, and the point charge are both placed at the center of the DPD particle.

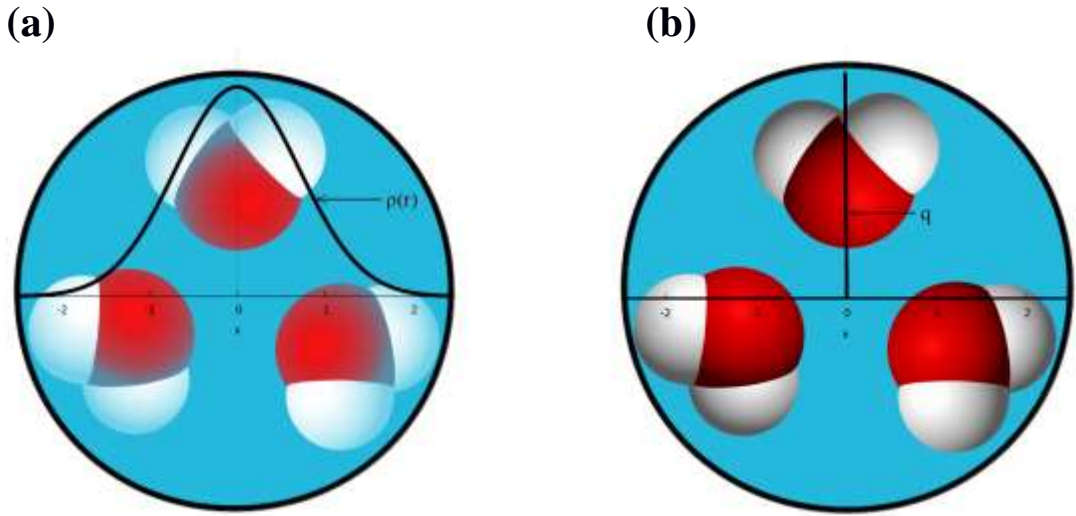

**Fig. 1**. (Color online) A schematic representation of the charge models considered in this work. (a) charge distribution model, $\rho(r)$, with its maximum at the center of mass of the DPD particle (blue filled circle). (b) Point charge model. The water molecules inside the DPD particles are meant to illustrate that the coarse graining degree is equal to three. Inspired from reference [35].

Let us start by comparing the structure of a fluid made up of a single polyelectrolyte in a solvent using point charges with that obtained from distributions of charge solved using the Ewald sums. Figure 2(a) shows the radial distribution functions for particles with opposite sign of the electric charge (those on the polyelectrolyte with the counter ions), and the radial distribution functions for the beads that make up the polyelectrolyte with each other, namely charges of equal sign are shown in Fig.2(b). The coarse – graining degree for all the cases shown in Fig. 2 is three, which is why the DPD interaction parameter is chosen as $a_{ij} = 78.0$.

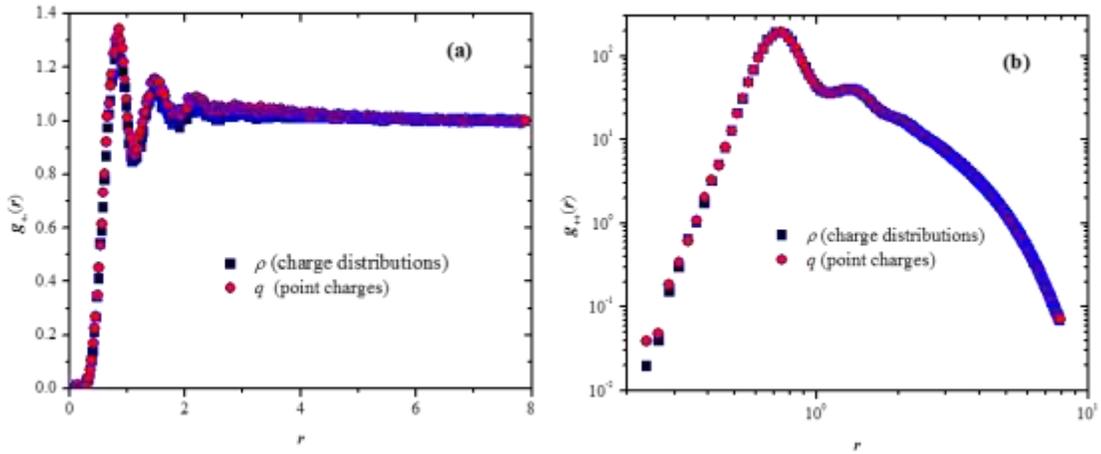

**Fig. 2** (Color online) (a) Radial distribution function for beads with charges of opposite sign, i. e., those on the polyelectrolyte, and the counter ions. (b) Radial distribution function for particles with charges of the same sign (++) for a linear polyelectrolyte with $N=32$ beads, $a_{ij}=78.0$ and total charge per bead equal to $q=0.4e$. The (blue) squares are the results obtained for charge distributions, while the (red) circles are the results obtained using point charges.

Clearly, the same structure is obtained for a strongly charged polyelectrolyte if one uses point charges (filled circles in Figs. 2(a) and (b)), or distributions of charge (solid squares in Figs. 2(a) and (b)), provided the DPD conservative interaction parameters are chosen strong enough to prevent the formation of ionic pairs. The structure of the pair distribution function between beads of the same sign (Fig. 2(b)) is larger due to the fact that those beads are joined by harmonic springs, forming the polyelectrolyte. Hence, one has to determine what constitutes a strong enough DPD interaction parameter and exactly under what conditions point charges can be used to replace charge distributions. With that end in mind, calculations were carried out of the radial distribution function for a system of simple, monomeric electrolytes and their counter ions with point charges, keeping constant the strength of their DPD conservative interaction while increasing the value of the point charges within the beads. The results are presented in Fig. 3 for three different values of the charge included into the DPD beads that make up the electrolytes in the solution.

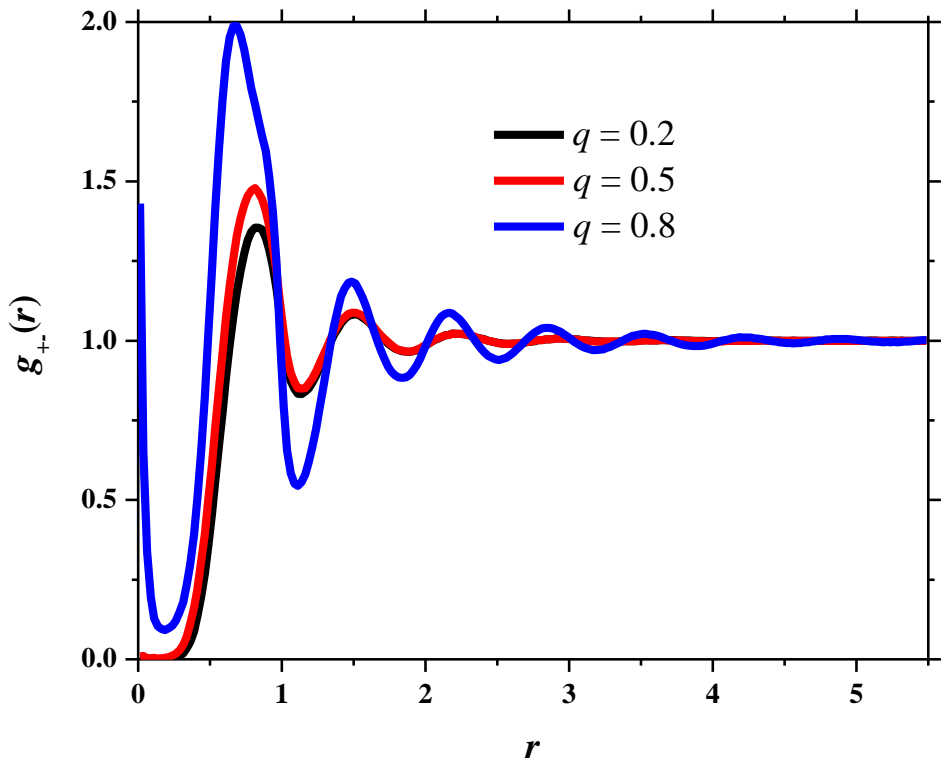

**Fig. 3** (Color online) Radial distribution function of monomeric electrolytes (+ -) immersed in a neutral solvent, with $a_{ij} = 78$, using the model of point charges with values of $q = 0.2e$ (black line), $q = 0.5e$ (red line) and $q = 0.8e$ (blue line). The coarse – graining degree in all cases was set at $N_m = 3$.

When the strength of the point charges attached to the DPD beads is increased, the structure displayed by the pair distribution function is also increased, until ionic pairs start to form for $q = 0.8e$, see the blue line in Fig. 3. One notices that for values of the charge up to $0.5e$ the system remains stable. The periodicity of the exponentially decaying oscillations seen in Fig.

3 remains essentially constant because all particles in the fluid have the same diameter [36]. The point charge model breaks down of course once the charge is strong enough so that the electrostatic attraction overcomes the DPD short range repulsion, which for the cases presented in Fig.3 corresponds to the case where the point charges on the electrolytes are $q = 0.8e$. Since increasing the coarse – graining degree increases the strength of the non – electrostatic, DPD repulsion parameter, it is to be expected that a fluid with charges can be made stable for a given strength of the point charge by increasing the coarse – graining degree. To test this hypothesis we carried out simulations of monomeric electrolytes, their counter ions and the solvent, at increasing values of the intensity of the point charges on the DPD beads, while increasing also the intensity of the conservative repulsive DPD force parameter, $a_{ij}$. The results of these calculations are shown in Fig. 4.

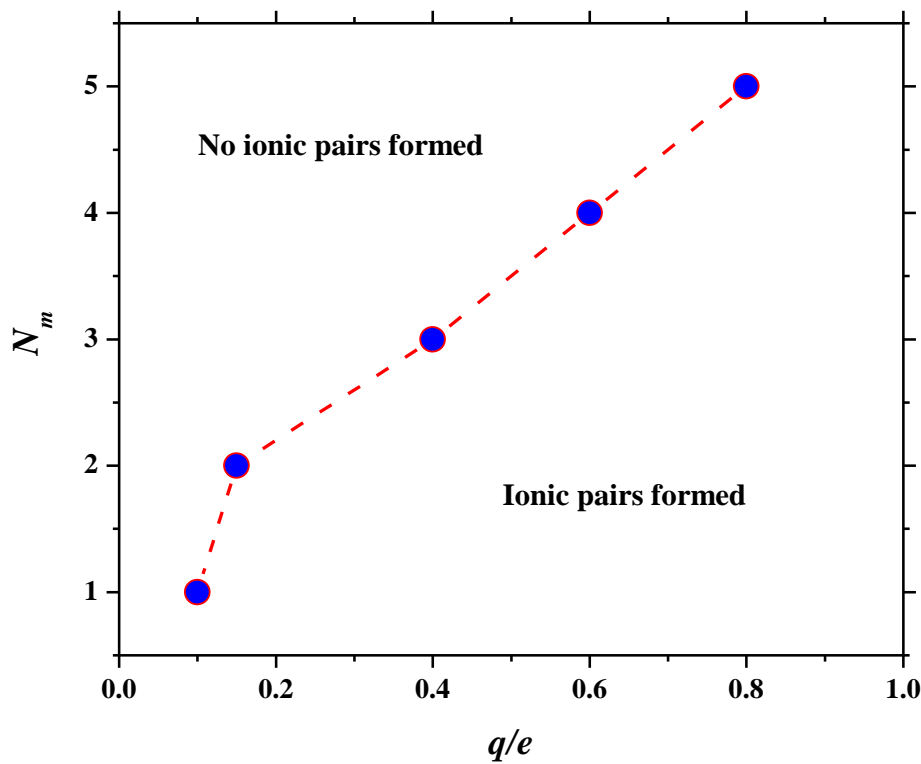

**Fig. 4** (Color online) Ionic pair formation diagram obtained when the charge on monomeric electrolytes and the coarse – graining degree ($N_m$) are increased. The dashed red line indicates the border between the phase where the system remains stable, i.e., no ionic pairs are formed (above the line) and where pairs of ions with opposite charge are formed (below the line). See Section III for details.

The criterion used to determine when ions of opposite charge formed an ionic pair in Fig. 4 was the following. If the relative distance between those particles was smaller than $0.7R_c/2$, it was considered that those particles formed an ionic pair, since the first peak found in the radial distribution function of neutral particles is at $0.7R_c$ [33]. As expected, increasing the coarse – graining degree ($N_m$), which is tantamount to increasing $a_{ij}$, translates into a stronger repulsion between DPD beads and prevents the formation of ionic pairs. We have constructed a three – dimensional diagram that helps determine what coarse – graining degree should be used to avoid the formation of ionic pairs when using point charges, which is presented in Fig. 5.

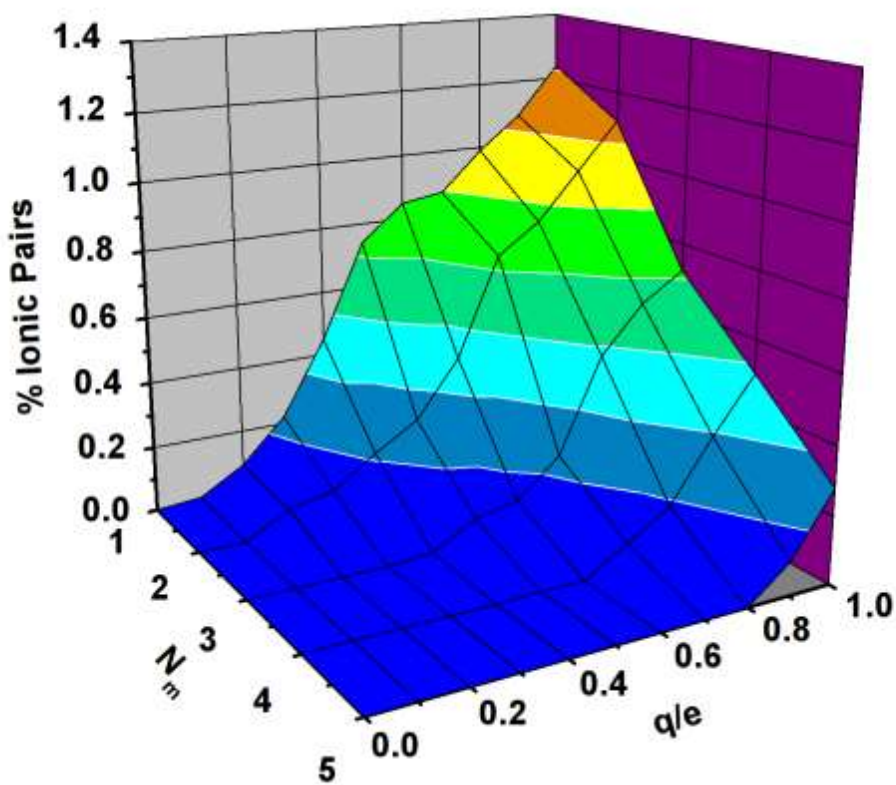

**Fig. 5** (Color online) Three - dimensional diagram corresponding to the formation of artificial ionic pairs, where the percentage of pairs formed during the simulation is shown as a function of the coarse – graining degree $N_m$ and the value of the point charge $q$.

There is ample range of phase space one can choose to avoid the formation of ionic pairs if point charges are used in conjunction with appropriate DPD interaction parameters, as Fig. 5 shows, and this fact can be taken advantage of when performing simulations of polyelectrolytes in solution, for example. Evidently, for very strongly charged polyelectrolytes the value of the repulsive DPD interaction parameter must be raised also, but for mesoscopic scale simulations this actually becomes a more efficient approach because the number of atom or molecules enclosed into DPD beads becomes larger [37].

Lastly, we perform an additional test of the use of point charges in DPD by predicting the scaling exponent of the gyration radius, $R_g$, as a function of the polymerization degree for polyelectrolytes in solution under increasing ionic strength. It is well known [32] that for neutral polymers the radius of gyration scales with the polymerization degree *N* (not to be confused here with the coarse – graining degree, $N_m$) as $R_g \sim N^\nu$, where $\nu$ is the scaling exponent, and it is known to depend on the quality of the solvent, and the dimensionality of the system. For three – dimensional polymers under theta conditions $\nu = 0.5$, while for polymers in good solvent $\nu = 0.588$ [38]. The values for polyelectrolytes are not well known, as the arguments for the scaling hypothesis are not supposed to hold for long range interactions, such as the Coulomb interaction. However, there are works that have shown that a scaling relation between the polymerization degree and the gyration radius does exist [39 – 41], therefore, it is important to determine whether such scaling property can be obtained using DPD particles with point charges. As Fig. 6 shows, it is in fact possible to find scaling of the radius of gyration for polyelectrolytes under different solvent conditions as the ionic strength is increased.

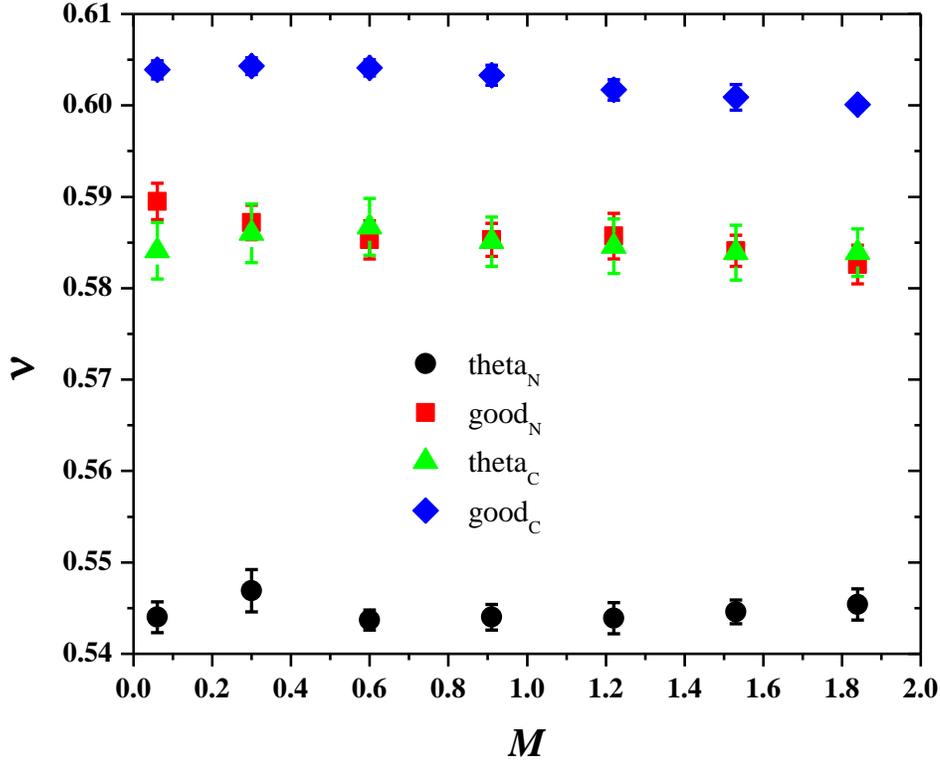

**Fig. 6** (Color online) Scaling exponent $\nu$ obtained for a neutral polymer in a theta solvent (black circles), for a neutral polymer in good solvent (red squares), for a polyelectrolyte in a theta solvent (green triangles), and for a polyelectrolyte under good solvent conditions (blue rhombi), at increasing values of the tetravalent salt concentration, $M$. Point charges where used for all calculations, with values of $q = 0.4e$, and $a_{ij}$=78. See Section III for details.

To obtain the data in Fig. 6 extensive simulations of neutral polymers and polyelectrolytes were carried out at several values of the polymerization degree, to obtain a scaling exponent for each ionic strength (see details in Section III). The reason for including neutral polymers is to provide an additional test of our approach, since even in that case electrostatic interactions with point charges are included in the fluid, to vary the ionic strength. The average value of $\nu$ obtained for a neutral polymer is a solution with increasing ionic strength under theta – solvent conditions is found to be $\nu = 0.545$, see the black circles in Fig. 6. This prediction is reasonably close to the value obtained for polymers in theta solvents in electrically neutral systems ($v = 0.5$) [32]. The fact that our prediction is slightly larger than 0.5 indicates that the salt ions associate with the polymer and impart it with a somewhat larger persistence length due to electrostatic repulsion [40]. Figure 6 shows also that the scaling exponent of the gyration radius is essentially the same for a fully charged polyelectrolyte in a theta solvent, and for a neutral polymer under good solvent conditions. The well – known prediction for the scaling exponent of the radius of gyration of polymers in good solvent, $v = 0.588$ [38], is very well reproduced when the polymer is subject to increasing ionic strength (red squares in Fig. 6). Additionally, the scaling of the polyelectrolyte under theta conditions is the same as if it was a neutral polymer in a good solvent, i. e., the quality of the solvent can be modified with the inclusion of the electrostatic interactions. When the ionic concentration is increased, the scaling exponent of the polyelectrolyte in a theta solvent remains more or less constant and equal, within statistical uncertainty, with respect to the scaling exponent of the neutral polymer under good solvent conditions. Lastly, the polyelectrolyte under good solvent conditions has a scaling exponent close to Flory's prediction for three – dimensional polymers in good solvent, $v = 3/5$ [32]. This is a weakly charged polyelectrolyte, which is why its scaling exponent is smaller than that obtained for strongly charged polyelectrolytes [39]; this prediction is consistent with values obtained experimentally for single – stranded DNA molecules [40]. It should be remarked that the choice of tetravalent salt ions is made in this work following the results from simulations carried out with different models and methods [42], which in turn are inspired by experiments [43].

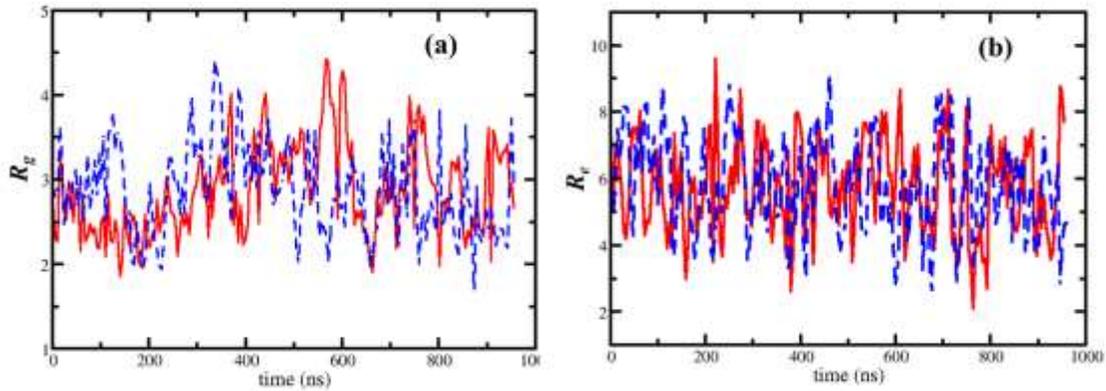

**Fig. 7** (Color online) Comparison of the radius of gyration ($R_g$, Fig. 7(a)) and end – to – end distance ($R_e$, Fig. 7(b)) of the polyelectrolyte whose g(r) were presented in Fig. 2, with polymerization degree $N = 32$, obtained with point charges (red, solid line), and charge distributions (blue, dashed line) along 1000 ns of simulation time. The charge on each monomer of the polyelectrolyte was set to $q = 0.4e$. The time averaged value of $R_g$ obtained for both approaches is $(2.9 \pm 0.5)R_C$. For the end – to – end distance, the time averaged $R_e$ is $(6.0 \pm 1.4)R_C$ for point charges, and $R_e = (6.0 \pm 1.5)R_C$ for charge distributions. The coarse – graining degree used for these simulations was $N_m = 3$, i.e., $a_{ij} = 78$ in reduced units.

As an additional example of the robustness of the point – charge approach in DPD we present in Fig. 7 the comparison between the radius of gyration and the end – to end distance of one polyelectrolyte chain in the solvent, obtained with distributions of charge (dashed lines) and point charges (solid lines). The polyelectrolyte is the same as that whose radial distributions functions were presented in Fig. 2. It is clear that both methods yield the same results within the statistical noise. In fact, after performing the time average we obtain $R_g = (2.9 \pm 0.5)R_C$ for both approaches, while the end – to – end distance is $R_e = (6.0 \pm 1.4)R_C$ for point charges and $R_e = (6.0 \pm 1.5)R_C$ for charge distributions.

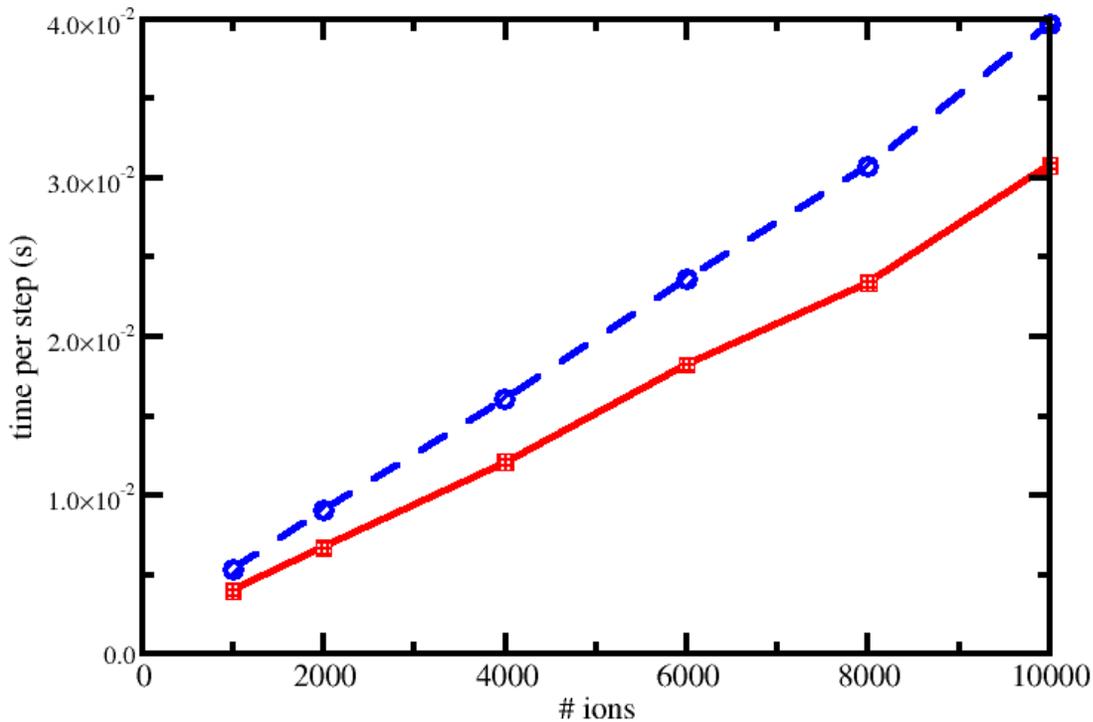

**Fig. 8** (Color online) Comparison of the numerical performance between point charges – solid line (red) - and charge distributions – dashed line (blue), tested on Nvidia GeForce GTX 980 graphics cards. The simulation box contains monomeric ions ($q = 0.4e$) and their counter ions ($q = - 0.4e$), and solvent particles. The conservative DPD repulsion constant for all species is $a_{ij} = 78$ (reduced units).

An additional advantageous aspect of the point – charge method is that it is computationally more efficient than the charge distribution method, as Fig. 8 shows. In it, the numerical performance of a system composed of monomeric ions, their counter ions and solvent particles is compared for the two methods. One sees that the time per step taken in the discreet integration of the equation of motion is smaller for the point charge approach, especially as the number of charges in the system is increased, where the savings in time can be ~ 25 %. Both systems were run on the same platform, namely Nvidia GeForce GTX 980 graphics cards. Six different sets of ions were modeled, with their counter ions, namely, 500, 1000, 2000, 3000, 4000 and 5000. For these simulations, the cubic cell simulation had sides equal to $L = $ 10.00, 12.59, 15.87, 18.17, 20.00 and 21.54 respectively, in reduced DPD units. The charge for the ions was set at $q = 0.4e$ while the charge of counter ions was $q = -0.4e$; the DPD repulsive interaction for all species was $a_{ij} = 78.0$. The rest of the simulation parameters remained unchanged, see Section III for additional details.

## V. CONCLUSIONS

The most commonly used method to include long – range interactions, such as Coulomb interactions in numerical simulations is still Ewald sums. On the other hand, DPD has proved to be a very useful tool to predict equilibrium and dynamic phenomena in soft matter systems. The incorporation of Ewald sums into DPD has traditionally used distributions of charge placed into the DPD beads, because it was originally thought that the soft nature of the DPD interactions could lead to the formation of ionic pairs of opposite charge, if point charges were used, which might lead to a singularity in the electrostatic interaction. However, this is true only for the smallest coarse – graining degree, which renders the mesoscopic reach of the DPD model inefficient. Here we have shown that for larger values of the coarse – graining degree, where DPD is truly mesoscopic, such artificial formation of ionic pairs with point charges can be avoided, and one obtains correct predictions. This means that one can use the full machinery of the optimized versions of the Ewald sums [12, 13] into the DPD methodology to improve the efficiency of the numerical simulations of increasing size, for applications to increasingly complex phenomena. Of course, one can continue using charge distributions in DPD using either Groot's method [16], or Ewald sums with the help of eq.

(12), and doing so should lead to reliable simulations, but for relatively large values of the coarse – graining degree this is not necessary and point charges can be used instead.

**Notes**

The authors declare no competing financial interest.

**Acknowledgments**

This work was supported by SIyEA-UAEM (Projects 3585/2014/CIA and 3831/2014/CIA). KATM thanks CONACyT, for a graduate student scholarship. All simulations reported in this work were performed at the Supercomputer OLINKA located at the Laboratorio de Bioingeniería Molecular a Multiescala, at the Universidad Autónoma del Estado de México. The authors are grateful to *Red Temática de Venómica Computacional y Bioingeniería Molecular a Multiescala.*